\documentclass[manuscript]{aastex}
\usepackage{latexsym,epsfig,graphics,graphicx,fullpage,amsmath}
\usepackage{color}
\def \be{\begin{equation}}
\def \ee{\end{equation}}
\def \bea{\begin{eqnarray}}
\def \eea{\end{eqnarray}}
\def \etal{{et al.\ }}

\begin{document}

\title{Probing Primordial Magnetic Fields Using Ly$\alpha$ Clouds}
\author{Kanhaiya L. Pandey\altaffilmark{1} AND Shiv K. Sethi\altaffilmark{1}}
\altaffiltext{1}{Raman Research Institute, Bangalore 560080, India}

\begin{abstract}
From previous studies of the effect of primordial magnetic fields on early structure 
formation, we know that the presence of primordial magnetic fields during early 
structure formation could induce more perturbations at small scales (at present 1--10 
$h^{-1}$Mpc) as compared to the usual $\Lambda$CDM theory. Matter
 power spectrum 
over these scales are effectively probed by cosmological observables
such as shear correlation  and Ly$\alpha$ clouds,  In this paper we 
discuss the  implications  of primordial magnetic fields 
on the distribution of Ly$\alpha$ clouds. We simulate 
the line of sight density fluctuation including the contribution coming from 
the primordial magnetic fields.
We compute the evolution of Ly$\alpha$ opacity for this case and compare our 
theoretical estimates of Ly$\alpha$ opacity with the existing data to  
constrain the parameters of the primordial magnetic fields. We also discuss the case when the two density fields are correlated. Our analysis yields 
an upper bounds of roughly $0.3\hbox{--}0.6 \, \rm nG$ on the magnetic field strength for  a range of nearly scale invariant models, corresponding to magnetic 
field power spectrum index $n \simeq -3$. 
\end{abstract}


\section{Introduction}

In the past 10 years, cosmological weak lensing and the study of 
 Ly$\alpha$ clouds 
in the redshift range
 $2 \la  z \la  5$ have emerged  as reliable  methods to 
 precisely determine the   matter power spectrum on 
 scales below $10 \, \rm h^{-1} Mpc$. In particular, 
these  methods  can  estimate the matter power spectrum at small scales which
are not directly accessible to other methods e.g. galaxy surveys (for details
and further references see e.g. \citet{Munshi08,Hoekstra08,Croft98,Croft99,Croft02}).

Large scale magnetic fields  been observed in galaxies and clusters of
 galaxies with the coherence lengths up to 
$\simeq$ 10\hbox{--}100 kpc (for a review see e.g. \citet{Widrow02}). There is
 also  some  evidence of coherent magnetic fields up to super-cluster 
scales \citep{Kim89}.
These  fields play an important role in various astrophysical processes.
Still little  is known about the origin of 
 large scale cosmic magnetic fields,  and their role in the evolutionary history of the universe. These
fields could have originated from  dynamo amplification of very tiny
 seed magnetic fields $\simeq 10^{-20} \, \rm G$ (e.g  \citet{Parker79,ZRS83,RSS88}). It is also possible 
that much larger primordial magnetic fields ($\simeq 10^{-9} \, \rm G$)  were generated during the 
inflationary phase \citep{TW88,Ratra92} and the large scale magnetic field
observed today are the relics of these fields.   These  large 
scale primordial fields  could be directly detected by upcoming  and future
radio interferometers such as LOFAR and SKA (for details see e.g. 
{\tt www.lofar.org} and {\tt www.skatelescope.org/pages/page\_sciencegen.htm}). 

The impact of large scale primordial magnetic fields on CMBR temperature 
and polarization anisotropies has been studied in detail (e.g. \citet{Kandu98b, 
Kandu02,Sesh01,Mack02,Lewis04,GS05,TS06,Sethi05,SBK08,Sethi09,Sethi10,Tina05,
Giovannini08,Yamazaki08,Sesh09,Trivedi10,Trivedi12}). 
\citet{Wasserman78} demonstrated  that primordial magnetic fields can
induce density perturbations in the post-recombination universe. Further
 work along these lines
have investigated the impact of this effect for the formation of first structures, 
reionization of the universe, and the signal from redshifted HI line from the epoch of 
reionization \citep[e.g.][]{Kim96,GS03,Sethi05,TS06,sbk,Sethi09}. 
 The matter power spectrum induced by
primordial magnetic fields can dominate the matter power spectrum of  
the standard $\Lambda$CDM model at small scales. Probes
such as cosmological  weak gravitational lensing 
can directly probe this difference and therefore reveal the presence of primordial fields 
and put additional constraint on their strength \citet{PS12}.

In this paper we attempt to constrain primordial magnetic 
fields  within the framework of the distribution of Ly$\alpha$ clouds
 in the IGM in the redshift range $2 \la z \la 5$. These clouds have been shown to originate in the mildly
non-linear density regions of the IGM (\citet{Cen94}). 
This has allowed development of  detailed semi-analytic methods 
to understand the observed
properties of these clouds \citep[e.g.][]{Bi95,Hui97,CTR01A, CTR01B}. 
Adopting a semi-analytic approach, we simulate 
density fluctuation along the line of sight, including the contribution 
of matter perturbations 
induced by these magnetic fields. We compute effective Ly$\alpha$ opacity of the IGM for this 
computed Ly$\alpha$ cloud distribution and 
 compare our results with the existing data  \citep[e.g.][]{FG08}.

In the previous analyses,  
the density perturbations induced by  magnetic fields 
are assumed to be uncorrelated to the density field generated by 
the usual $\Lambda$CDM model. Recently, \citet{Caldwell11} showed that if the conformal invariance
 of electromagnetism is broken during the 
inflation and thus produced the primordial magnetic fields, these
 magnetic fields may be correlated 
with the primordial density perturbations. In our analysis
 we have made a separate case for such fields. 

Throughout the paper, we used flat (k=0) $\Lambda$CDM universe with $\Omega_{\rm m}$ = 0.24, 
$\Omega_{\rm b}$ = 0.044, $h$ = 0.73 and $\sigma_{\rm 8}$ = 0.77.
\\

\section{Primordial Magnetic Field And Induced Matter Power Spectrum}

In the primordial theory of the  magnetic fields, it is postulated that
 the large scale 
primordial magnetic fields of  strengths $\simeq 10^{-9} \, \rm G$ were present in the 
very early universe; these fields  could have originated during the inflation.
 They are assumed 
to be tangled magnetic fields which can be characterized by a power-law 
power spectrum:
$M(k) = A k^n$. In the prerecombination era, the magnetic fields
are dissipated at scales below a scale corresponding to 
$k_{\rm max} \simeq 200 \times (10^{-9} \, {\rm G}/B_{\rm eff})$ \citep[e.g.]{jedam98,kandu98a}. 
Here $B_{\rm eff}$ is the RMS at this cut-off
scale for a given value of the spectral index, $n$. Another possible normalization, commonly used in the
 literature, is the value of RMS at $k = 1 \, \rm Mpc^{-1}$, $B_0$. 
These two normalizations are related as: $B_{\rm eff} = B_0 k_{\rm max}^{(n+3)/2}$. 
It is possible to present results using either of the pair
 $\{B_{\rm eff},n\}$ or  $\{B_0,n\}$.

The  PMF  induced matter perturbations 
 grow in the post recombination era by gravitational instability.
 The matter 
power spectrum of these perturbations is given by: $P(k) \propto k^{2n+7}$, 
for $n < -1.5$, the range of spectral indices we consider here \citep{Wasserman78,Kim96,GS03}. 

Magnetic field induced matter perturbations can only grow for scales above the 
magnetic field Jean's length: $k_{\rm J} \simeq 15 \times (10^{-9} \, \rm G/B_{\rm eff})$ \citep[e.g.][]{Kim96,Tina2011}.
The dissipation of tangled magnetic field in the post-recombination era also
results in an increase in the thermal Jeans length \citep{Sethi05,SBK08}. For 
most of the range of magnetic field strengths and the physical
setting (Lyman-$\alpha$ clouds at a temperature of $\simeq 10^4 \, \rm K$)
 considered here,
the  scale corresponding to $k_{\rm J}$ generally are comparable to or smaller
than  the  thermal Jeans length. 

For our computation, we need to know the  time evolution of the matter power
spectrum induced by tangled magnetic fields. It can be shown that the dominant
 growing mode in this case  has the same time dependence as the $\Lambda$CDM model 
(see e.g. \citet{GS03} and references therein)

\section{The Simulation: Density Fluctuation Along The Line of Sight: Distribution of Ly$\alpha$ Clouds}

We describe a brief outline of the numerical simulation in this section.
 Hydrodynamical simulations 
have shown that Ly-$\alpha$ clouds are mildly non-linear 
($\delta \la 10$) regions of the IGM at high redshifts. This allows one 
to analytically derive important observables from the Lyman-$\alpha$ clouds 
semi-analytically, in terms of a few parameters denoting the 
ionization, thermal, and dynamical state of the clouds. 

Here we have closely followed the semi-analytic prescription 
given in \citet{BD97}. In this paper we have considered two cases of primordial magnetic 
field induced matter perturbations : (a) pure $\Lambda$CDM matter perturbations 
and primordial magnetic field (PMF) induced matter perturbations are uncorrelated 
(b) those two are correlated. 
In both cases we compute two separate line of sight density 
 (and velocity) fields each corresponding to a single kind of matter 
perturbations. In the former  case, these fields are drawn from different
realizations and in the latter the fields are generated from the same realization. We add these two density (and velocity) 
fields to get the final density (and velocity) fields in the IGM.
To simulate line of sight IGM density and velocity fields for a given three dimensional matter power spectrum (inflationary/PMF 
induced), first we calculate the corresponding three dimensional baryon power spectrum, which corresponds to the original 
power spectrum smoothed over the scales below the larger 
of the thermal or Magnetic  Jeans scale $x_b$
\be
P^{(3)}_{\rm B} (k,z) = \frac{P^{(3)}_{\rm DM}(k,z)}{[1 + x^2_b(z)k^2]}
\ee
where
\be
x_{b}(z) = \frac{1}{H_0} \left[ \frac {2 \gamma k_{\scriptscriptstyle \rm B} T_{\rm m} (z)}{3 \mu m_{p} \Omega_{\rm m} (1+z)} \right]^{1/2}
\ee
then we compute one dimensional baryon (density, velocity and  density-velocity) power spectra, which will be used in the further 
computation. We note here that the relevant scale of smoothing for 
the range of magnetic field values and the IGM temperatures we consider 
is thermal Jeans scale and not the magnetic Jeans scale.  The one-dimensional power spectra can be computed using the following relations
\be
P^{(1)}_{\rm B} (k,z) = \frac {1}{2 \pi} \int^\infty_{\lvert k \rvert} dk' k' P^{(3)}_{\rm B} (k,z)
\ee
\be
P^{(1)}_{\rm v} (k,z) = \dot{a}^2(z)k^2 \frac {1}{2 \pi} \int^\infty_{\lvert k \rvert} \frac{dk'}{k'^3} P^{(3)}_{\rm B} (k,z)
\ee
\be
P^{(1)}_{\rm Bv} (k,z) = i\dot{a}(z)k \frac {1}{2 \pi} \int^\infty_{\lvert k \rvert} \frac{dk'}{k'} P^{(3)}_{\rm B} (k,z)
\ee
where $a$ is the scale factor. 

The density ($\delta_0(k,z)$) and velocity ($v(k,z)$) fields in one dimension are two correlated Gaussian random fields 
(the correlation is given by the density-velocity power spectrum), we use inverse Gram-Schmidt procedure to simulate them 
in terms of two independent Gaussian random fields $w(k)$ and $u(k)$ of power spectra respectively $P_{\rm w}(k)$ and 
$P_{\rm u}(k)$ 

\be
\delta_0 (k,z) = D(z) [u(k) + w(k)]
\ee

\be
v (k,z) = F(z) i \dot{a} k \beta(k,z) w(k,z)
\ee
where D(z) and F(z) are the linear density and velocity growth factors. Functions $\beta(k)$, $P_{\rm w}(k)$ and 
$P_{\rm u}(k)$ are function of $P^{(3)}_{\rm B}(k)$, 

\be
\beta (k,z) = \int^\infty_{\lvert k \rvert} \frac {P^{(3)}_{\rm B}/k'^{3} dk'} {P^{(3)}_{\rm B}/k' dk'}
\ee
\be
P_{\rm w}(k) = \frac {1}{\beta(k)} \int^\infty_{\lvert k \rvert} \frac {P^{(3)}_{\rm B}(k')}{k'} dk'
\ee
\be
P_{\rm u}(k) = \frac {1}{2 \pi} \int^\infty_{\lvert k \rvert} P^{(3)}_{\rm B}(k') k' dk' - P_{\rm w}(k)
\ee

We compute $\delta_0(k,z)$ and $v(k,z)$ for both kinds of perturbations separately, the corresponding $\delta_{\rm B}(x,z)$
and $v(x,z)$ is computed by using Fourier transforms. And then we add the contribution from both the kinds together 
($\delta_{\rm B}(x,z) = \delta^{\rm infl}_{\rm B}(x,z) + \delta^{\rm pmf}_{\rm B}(x,z)$ and $v(x,z) = v^{\rm infl}(x,z) + v^{\rm pmf}(x,z))$ to 
get the final combined line of sight density and velocity fields. To compute one dimensional density field for the pmf induced 
perturbations we use the three-dimensional  matter power spectrum 
 (e.g. Gopal \& Sethi, 2003);  for 
inflationary perturbations we use the standard $\Lambda$CDM power spectrum.

For our computations  we have generated the density and velocity  
 fields for 25 redshift bins between the redshifts 0 to 5. In each bin we have 
$\rm 2^{14}$ points resolving the Jeans scale by at least a 
factor of 4. The cutoff 
scale (Jeans scale, $x_b$)  is the larger of the thermal Jeans length and the magnetic Jeans length. 

To take into account the  non-linearity of density perturbations in 
the IGM we  use  lognormal distribution of the IGM density
 field \citet{BD97}, 
thus the number density of baryons in IGM is taken to be,

\be
n_{\scriptscriptstyle \rm B} (x,z) = A e^{\delta_{\rm B} (x,z)}
\ee
A is a constant which can be determined using following relation,
\be
\langle n_{\scriptscriptstyle \rm B} (x,z)\rangle \equiv n_0(z) = A \langle e^{\delta_B (x,z)}\rangle
\ee
since $\delta_{\rm B} (x,z)$ is a Gaussian random variable,
\be
\langle e^{\delta_{\scriptscriptstyle \rm B} (x,z)}\rangle = e^{\langle\delta_{\scriptscriptstyle \rm B}^2 (x,z)\rangle}
\ee
thus 
\be
n_{\scriptscriptstyle \rm B} (x,z) = n_0(z) {e^{(\delta_{\scriptscriptstyle \rm B} (x,z)-\langle\delta_{\scriptscriptstyle \rm B}^2(x,z)\rangle)}}
\ee
where $n_0(z)$ is the background baryon number density given by,
\be
n_0(z) = \frac {\Omega_{\rm B} \rho_c}{\mu_{\scriptscriptstyle \rm B} m_p} (1+z)^3
\ee

\section{Calculation of Ly$\alpha$ opacity}

The optical depth $\tau$ is given by 
\be
\tau(\nu) = \int n_{\rm HI} (t) \sigma_a \left(\frac {\nu}{a}\right) dt
\ee
where $n_{\rm HI}$ is number density of neutral hydrogen in the IGM, $\nu$ is the observed 
frequency, which is related to redshift z by 
$z \equiv (\nu_a / \nu) - 1$, $\nu_a$ is the Ly$\alpha$ frequency at rest. 
The absorption cross section $\sigma_a$ is given by, 
\be
\sigma_a = \frac {I_a}{b \surd \pi} V \left(\alpha, \frac {\nu-\nu_a}{b \nu_a} + \frac {v}{b} \right)
\ee
where parameter $b = (2 k T / m_p)^{1/2}$ is the velocity dispersion and $v(x)$ is the peculiar velocity field, 
$\alpha \equiv 2 \pi e^2 \nu_a / 3 m_e c^3 b = 4.8548 \times 10^{-8} / b $,
$I_{\alpha} = 4.45 \times 10^{-18} cm^{-2}$ and V is the Voigt function. 

The number density of neutral hydrogen, $n_{HI}$ can be computed by solving ionization equilibrium equation,

\be
n_{\rm HI} (x,z) = \frac {\alpha [T(x,z)] n_{\scriptscriptstyle \rm B} (x,z)}{\alpha [T(x,z)] + \Gamma_{ci}[T(x,z)] + J(z)/[\mu_e n_{\scriptscriptstyle \rm B}(x,z)]}
\ee
where $T(x,z)$ is given by $T(x,z) = T_0(z) [{n_{\scriptscriptstyle \rm B}(x,z)}/{n_0(z)}]^{\gamma - 1}$ where $T_0(z)$ is the 
temperature of the IGM at the mean density and $\gamma$ is the polytropic index for the IGM; $\gamma$ 
captures the dynamical state of the IGM gas which 
gives rise to the observed Lyman-$\alpha$ absorption.  These parameters 
 are likely to take 
values in the ranges 4000 $\lesssim T_0 \lesssim$ 15,000 K and 1.3 $\lesssim \gamma \lesssim$ 1.6 (Hui \& Gnedin, 1997).
$\alpha(T)$, $\Gamma_{ci}(T)$ and $J(z)$ are recombination rate, collisional ionization 
rate and photo ionization rates in the IGM. For temperature $T \simeq 10^4 K$, the combination of these effects yields 
(Croft \etal 1998),
\be
\tau(z) \propto n_{\scriptscriptstyle \rm B}^2 T^{-0.7} = A(n_{\scriptscriptstyle \rm B}/n_0)^{2-0.7(\gamma -1)} , \nonumber \\[2ex]
\ee
\be
A = 0.946 \left(\frac{1+z}{4}\right)^6 \left(\frac{\Omega_{\rm B} h^2}{0.0125}\right)^2 \left(\frac{T_0}{10^4 K}\right)^{-0.7}
          \left(\frac{J}{10^{12} s^{-1}} \right)^{-1} \left[\frac {H(z)}{H_0} \right]^{-1}
\ee

To compare with the data we have computed effective optical depth which is the
 observable quantity in the form of decrease in 
observed flux ($ F \propto e^{-\tau} $) and is given by,
\be
\tau_{\rm eff}(z) = -\log\left [\langle \exp({-\tau})\rangle \right]
\ee

 The data which we have used for  comparison with simulation results
has been obtained  using high resolution spectral 
 observations like HIRES, ESI and MIKE having FWHM in 
the range of $6\hbox{--}44 \, \rm km \, sec^{-1}$ (\citet{FG08}), 
which  resolve  the Jeans scales over the redshifts we are considering. 
Since we are also resolving Jeans scale in our simulation, we can directly 
compare our theoretical results with 
this data without taking into account  the scale dependence of 
$\tau_{\rm eff}$ in our analysis. 

 The mean opacity  $\langle\tau\rangle$ and the effective 
opacity are computed by  averaging  over 
all the realizations of $\tau$ for a given 
redshift bin. 

\begin{figure}
   \centerline{
   \includegraphics[scale=0.7,angle=-90]{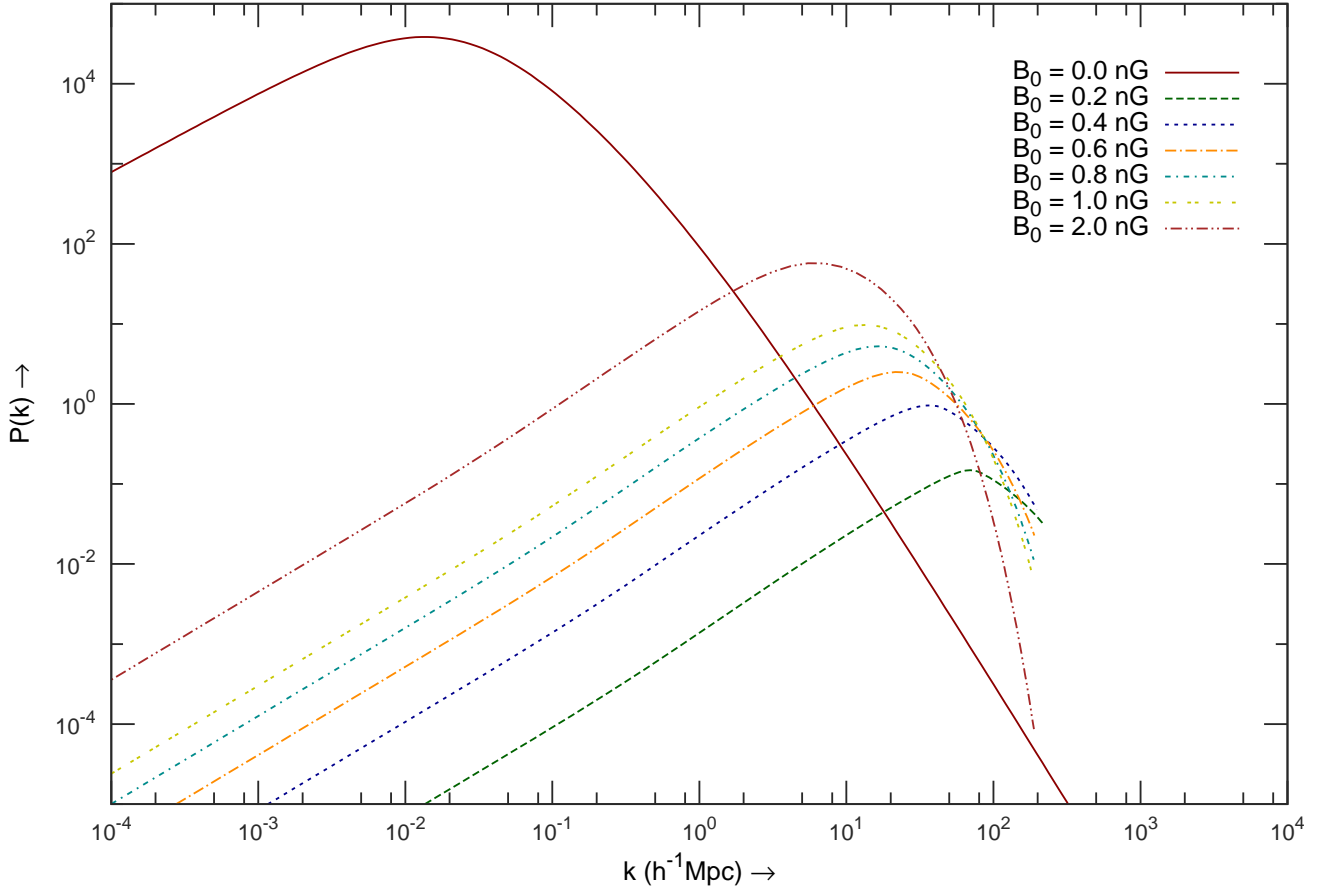}
}
   \caption[]{
The matter power spectrum for magnetic case, with added exponential cutoff and then smoothed around magnetic Jeans length $k_{\rm J}$, 
is displayed for various values of magnetic field strength $B_0$. Spectral index $n$ is -2.95 for each case. Along with that 
the red curve is matter power spectrum for pure $\Lambda$CDM non-magnetic case. 
}
\end{figure}
\vspace{0.1in}

\section{Results}
In Figure~1 we show the  matter power spectra at the present epoch 
for magnetic case, along with the pure $\Lambda$CDM (non-magnetic)
matter power spectrum, which has been used in our calculations, here an exponential cut-off around $k ~$ magnetic Jeans scale
is assumed. This figure shows that the magnetic field induced matter power spectra can dominate over the pure $\Lambda$CDM case
at small scales ($k \ge 1 \, \rm  h/Mpc$). The effect of this excess has
 already been studied in the context of early structure formation,
reionization, and weak-lensing signals (\citet{Sethi05,Sethi09}, \citet{PS12})
As an extension to that body of work   we explore the effect of this excess
on Ly$\alpha$ effective  opacity in this paper.

In Figure~2 we show the variation of Ly$\alpha$ opacity $\langle\tau\rangle$ 
with redshift for various values of magnetic field strengths.
The red dots with y-errorbars are the observed values of Ly$\alpha$ opacity $\tau_{\rm eff}$ (\citet{FG08}). It should be pointed out that the inclusion
of peculiar velocities in the computation of $\tau$ (Eq.~(17)) makes
a negligible difference to the value of either average or effective opacity.

Figure~1 corresponds to   the case when
matter perturbations induced by primordial magnetic fields and the
 inflationary matter perturbations are not correlated. 
The average opacity $\langle\tau\rangle$ is not an observable quantity.The aim of Figure~2 is to demonstrate that the inclusion of PMF 
matter perturbations enhances the average opacity of the IGM.

In Figure~3 we show the variation of $\tau_{\rm eff}$ with redshift for 
various values of magnetic field strength along with the observed evolution
of $\tau_{\rm eff}$. This plot is for the case when 
matter perturbations induced by primordial magnetic field and the inflationary matter perturbations are not correlated. 
Comparing this figure with Figure~2 we see that the slope of redshift
evolution of $\tau_{\rm eff}$ is far smaller than for  average opacity. This 
difference is owing to the fact that for HI column densities $N_{\rm HI} \ga 10^{14} \, \rm cm^{-2}$, the optical depth exceeds one. For column densities 
larger than this saturation value, the optical depth increases only as logarithm of the column density and therefore these clouds get a smaller weight in the 
the computation of $\tau_{\rm eff}$. As the average opacity of the IGM increases
sharply with increasing redshift (Eq.~(19)), this effect is more enhanced 
at higher redshifts. 

 A comparison between Figures~2 and~3 shows that an increase in $\langle \tau \rangle$ doesn't necessarily lead to an increase in 
$\tau_{\rm eff}$. In Figure~3 it is seen that $\tau_{\rm eff}$ is greater than 
the usual $\Lambda$CDM case for $z \la 3$ but falls below the predictions of this model for larger redshifts. 

We can understand this behaviour by the following set of arguments. The change
in the effective optical depth  
$d\tau_{\rm eff} \propto \sum_i \exp(-\tau_i) d\tau_i$, where $\tau_i$ refers to optical  depths 
of individual clouds. On the other hand, $d\langle \tau \rangle \propto \sum_i d\tau_i$. As seen in Figure~2, the inclusion of 
PMF density perturbations increase $\langle \tau \rangle$ or $ \sum_i d\tau_i > 0$, but $\sum_i \exp(-\tau_i) d\tau_i$ could be 
negative if $d\tau_i$ is negative wherever $\tau_i$ is smallest. 
 To elaborate this point,  In  Figure~6 
we have plotted the distribution of optical depths 
 $\tau_i$s for the 1.5~nG case ($z =4$), against the $d\tau_i =  \tau_i \ (2 \ nG) - \tau_i \ (0 \ nG); (z = 4)$.  It 
is clear from this figure that $\tau_i$ values are  small when  
 $d\tau_i$ is more negative, or this can make $\sum_i \exp(-\tau_i) d\tau_i$ 
negative, and thus it explains the decrease of $\tau_{\rm eff}$ 
even when there is increase in $\langle\tau\rangle$ with increasing magnetic field values.
 It should be pointed 
out that this behaviour of $\tau_{\rm eff}$ cannot be mimicked by a change in $J$, $\gamma$ (Eq.~(19)) or by a scaling 
of the power spectrum by changing the value of $\sigma_8$.

The Figure~4 and Figure~5 are for the same analysis as Figure~2 and Figure~3 respectively, but for the case when induced matter 
perturbations and inflationary matter perturbations are correlated. In Figure~4 the values of $\langle \tau \rangle $ are smaller in 
comparison to the corresponding values in the case of uncorrelated
 matter perturbations.

For detailed comparison with observations we performed the likelihood 
analysis for 
 the $\tau_{\rm eff}$ against the \citet{FG08} data as a function 
of four parameters, J ((1.4 to 2.0)10$^{-12}$), $\gamma$ (1.4 to 2.0), $B_0$ ((0.1 to 2.0) nG), and n (-2.80 to -2.99). To compute the posterior probability
for magnetic field parameters,  we marginalized the likelihood 
function  over the parameters J \& $\gamma$.
 Figure~7 shows the results of this  analysis
for the uncorrelated case. The curves from top to bottom are the contours for 5$\sigma$, 3$\sigma$ and 1$\sigma$ levels for 
a range of $\Delta\chi^2 = \chi^2_{\rm i} - \chi^2_{\rm min}$. We see that in this
 case for n = -2.90 the allowed values (by 5$\sigma$) 
of magnetic field are $B_0 <$ 0.6-0.7 nG, and for n = -2.95 is $B_0 < 1.3 \, \rm nG$.

In Figure~8, we compare this result with our previous analysis
 with the weak-lensing 
data \citet{PS12} and the present 
analysis with the correlated case: 
the lower triplet (red green and blue), solid and dashed
corresponds to the uncorrelated and the correlated cases respectively of the
 present analysis, 
whereas the upper triplet (dotted) correspond to our previous analysis with the weak-lensing data. It is clearly seen from Figure~8 that 
the constraints arising from the correlated case are not very different
from the uncorrelated case. Or the Lyman-$\alpha$ clouds do not provide
an appropriate physical setting for distinguishing  between these 
two cases. 
From Figure~8 it also  follows that our present constraints are 
considerably stronger that our previous analysis with the 
cosmological weak-lensing data.

\begin{figure}
   \centerline{
   \includegraphics[scale=0.7,angle=-90]{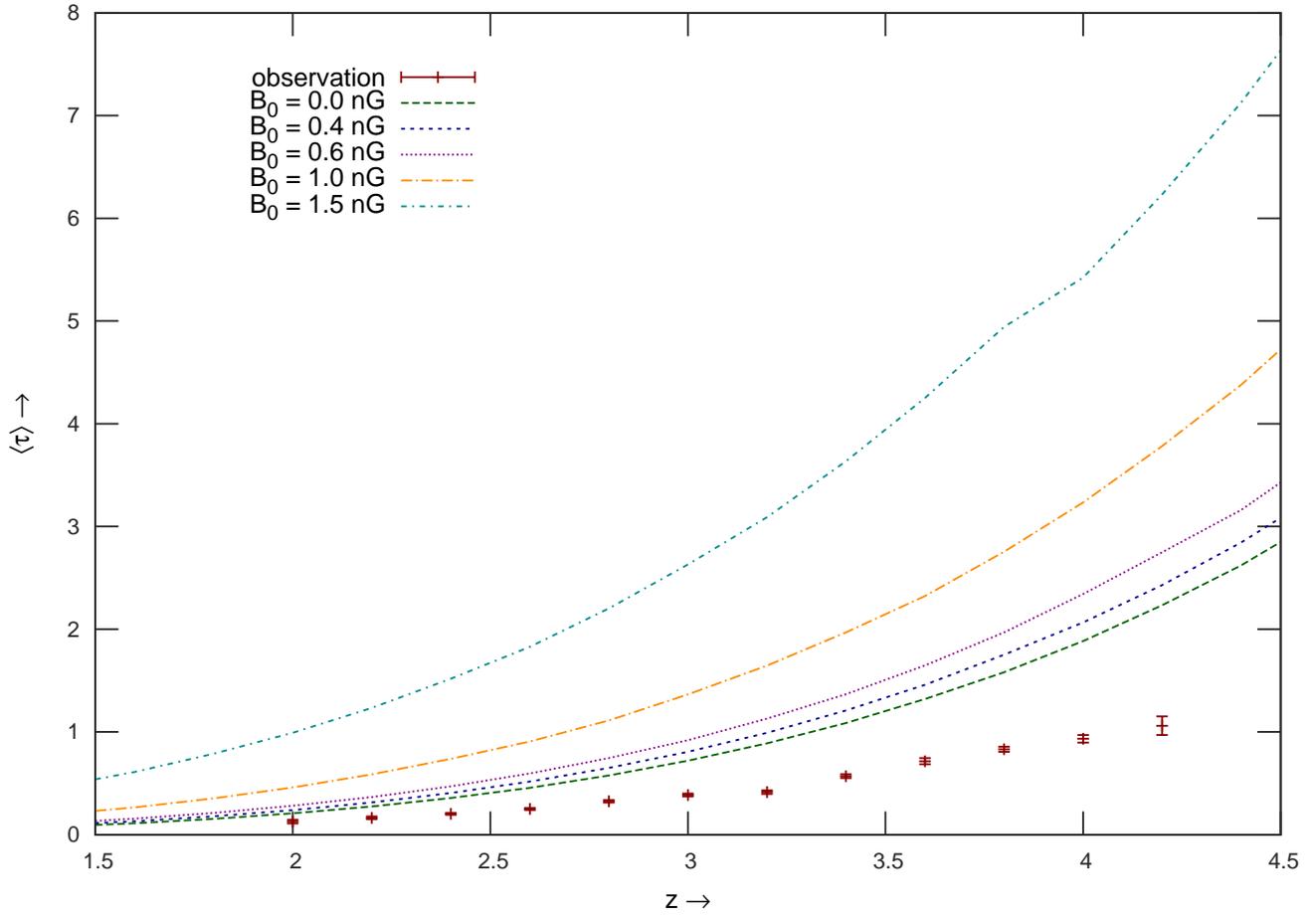}
}
   \caption[]{
The evolution of Ly$\alpha$ opacity $\langle\tau\rangle$ for the magnetic and non magnetic cases,
uncorrelated $\delta_{\rm infl}$ and $\delta_{\rm pmf}$ case.
}
\end{figure}
\vspace{0.1in}

\vspace{0.1in}
\begin{figure}
   \centerline{
   \includegraphics[scale=0.7,angle=-90]{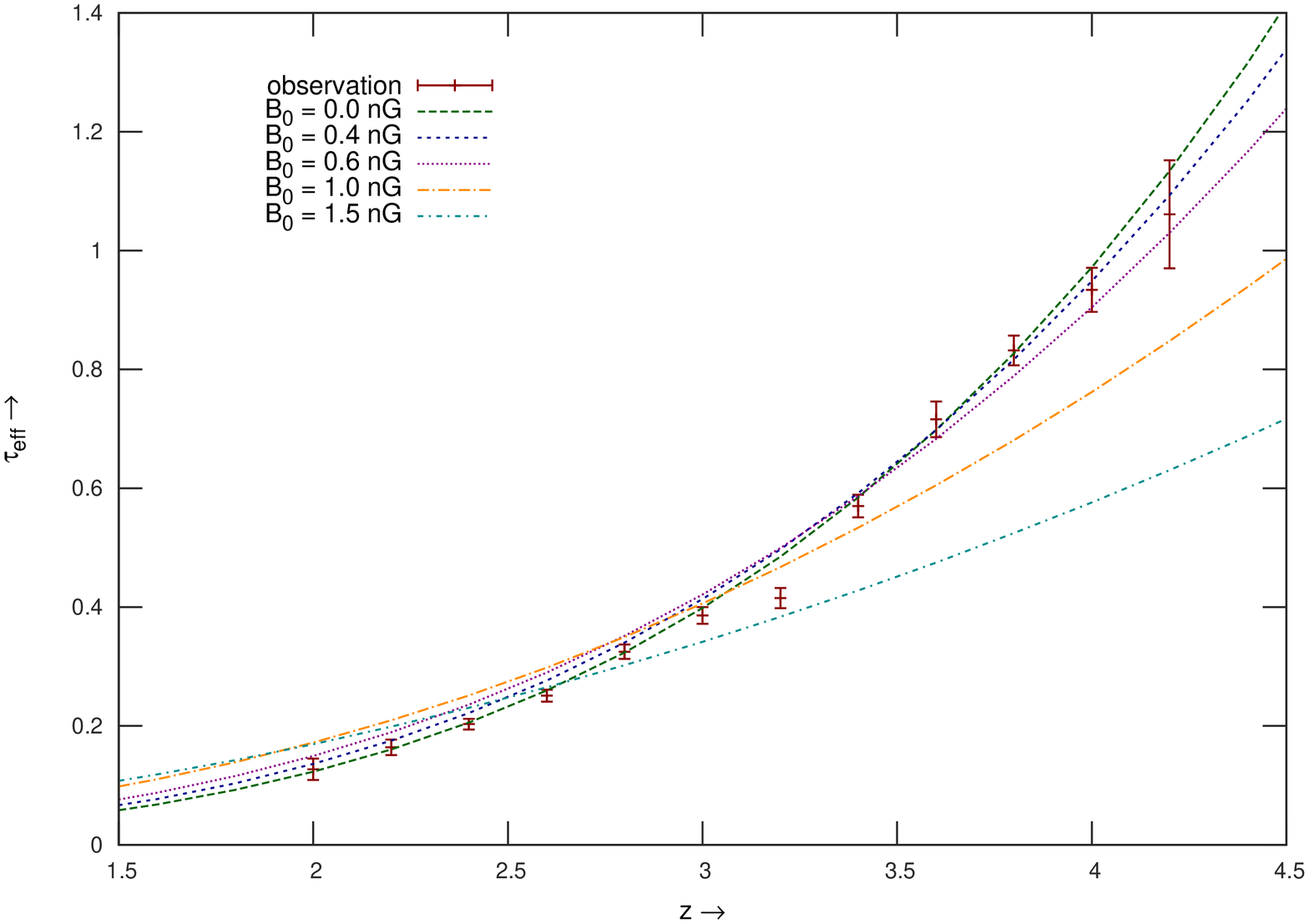}
}
   \caption[]{
The evolution of Ly$\alpha$ opacity $\tau_{\rm eff}$ for the magnetic and non magnetic cases,
uncorrelated $\delta_{\rm infl}$ and $\delta_{\rm pmf}$ case.
}
\end{figure}
\clearpage
\vspace{0.1in}
\begin{figure}
   \centerline{
   \includegraphics[scale=0.7,angle=-90]{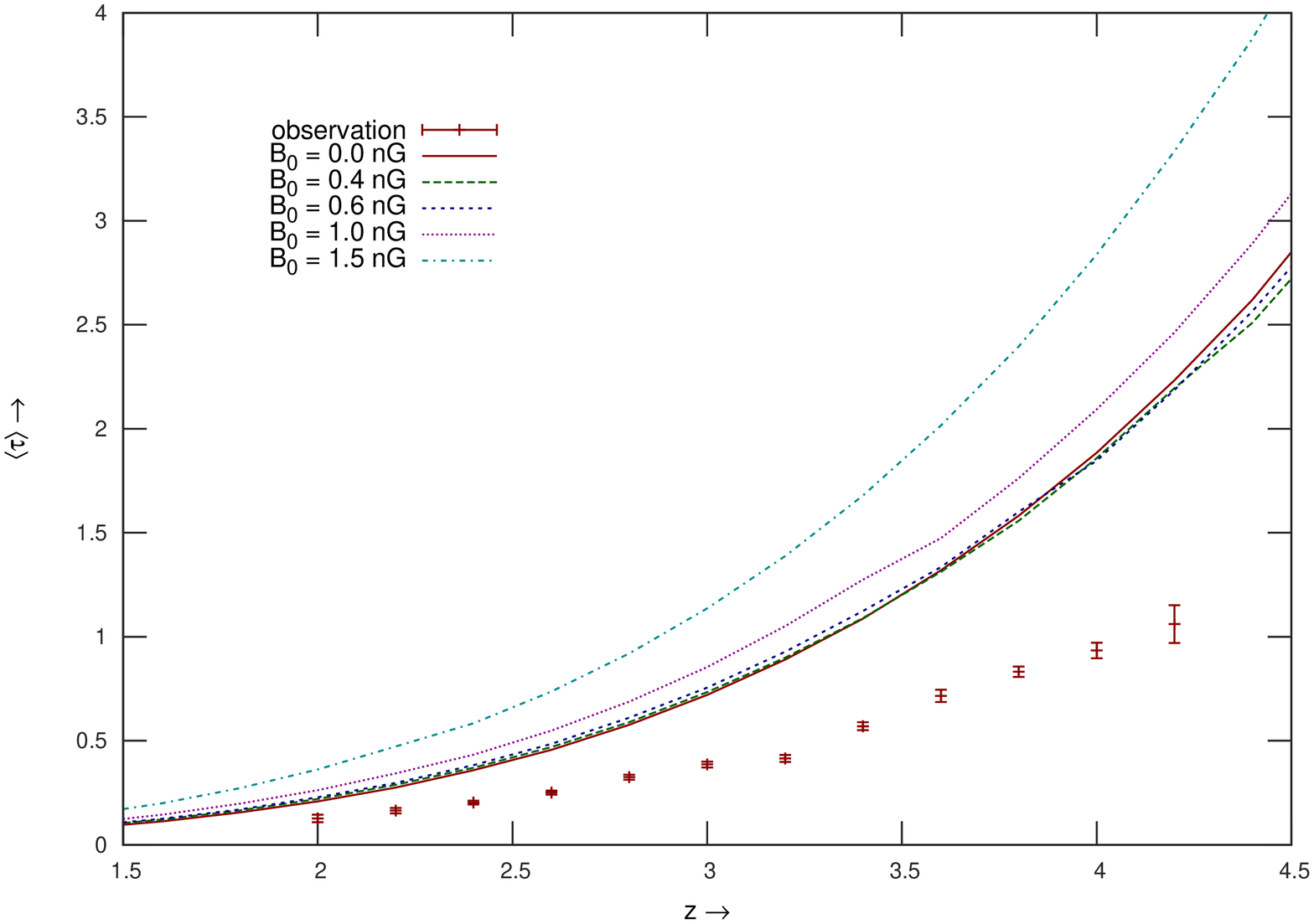}
}
   \caption[]{
The evolution of Ly$\alpha$ opacity $\langle\tau\rangle$ for the magnetic and non magnetic cases,
correlated $\delta_{\rm infl}$ and $\delta_{\rm pmf}$ case.
}
\end{figure}
\vspace{0.1in}

\vspace{0.1in}
\begin{figure}
   \centerline{
   \includegraphics[scale=0.7,angle=-90]{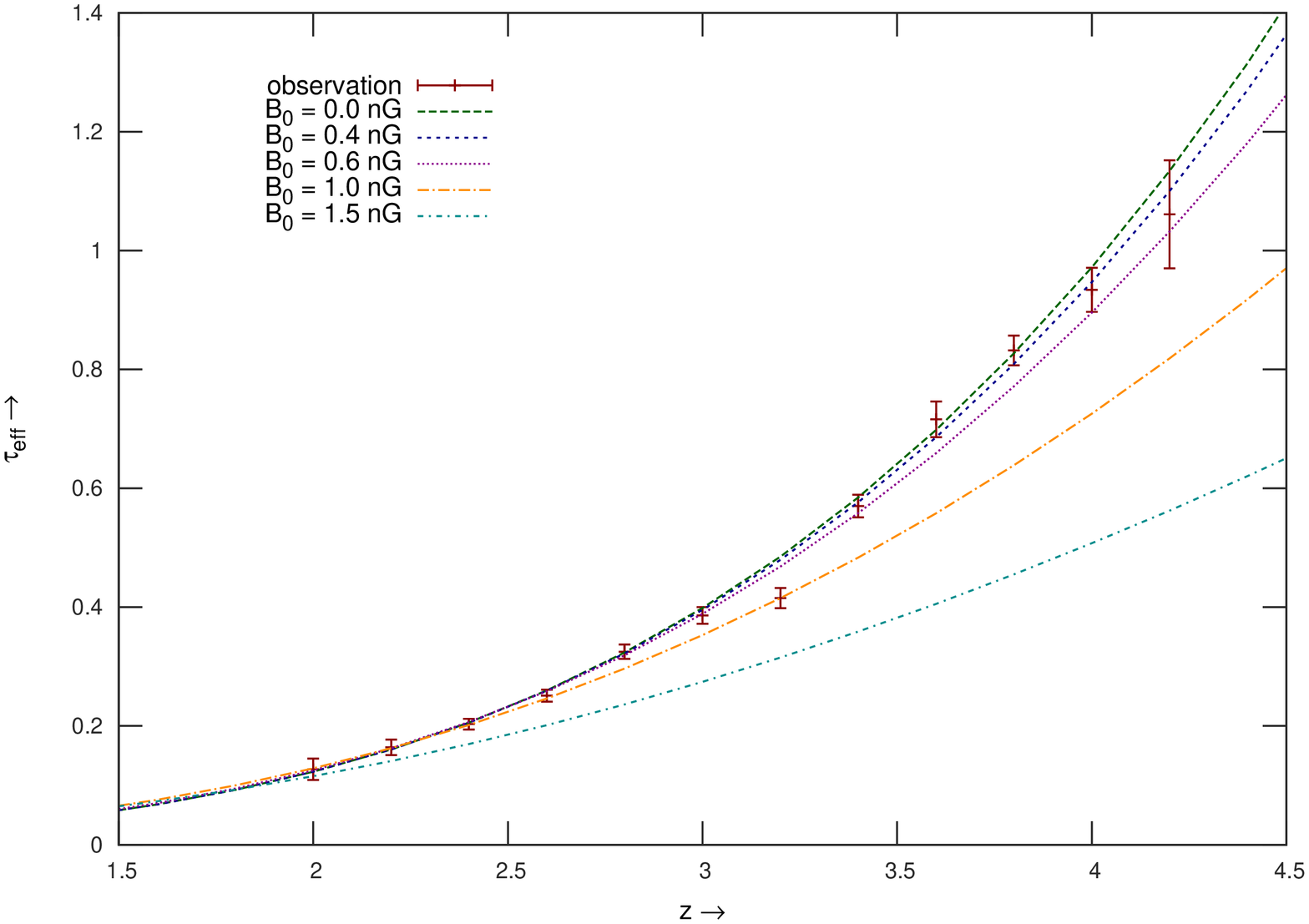}
}
\caption[]{
The evolution of Ly$\alpha$ opacity $\tau_{\rm eff}$ for the magnetic and non magnetic cases,
correlated $\delta_{\rm infl}$ and $\delta_{\rm pmf}$ case.
}
\end{figure}

\vspace{0.1in}
\begin{figure}
   \centerline{
   \includegraphics[scale=0.7,angle=-90]{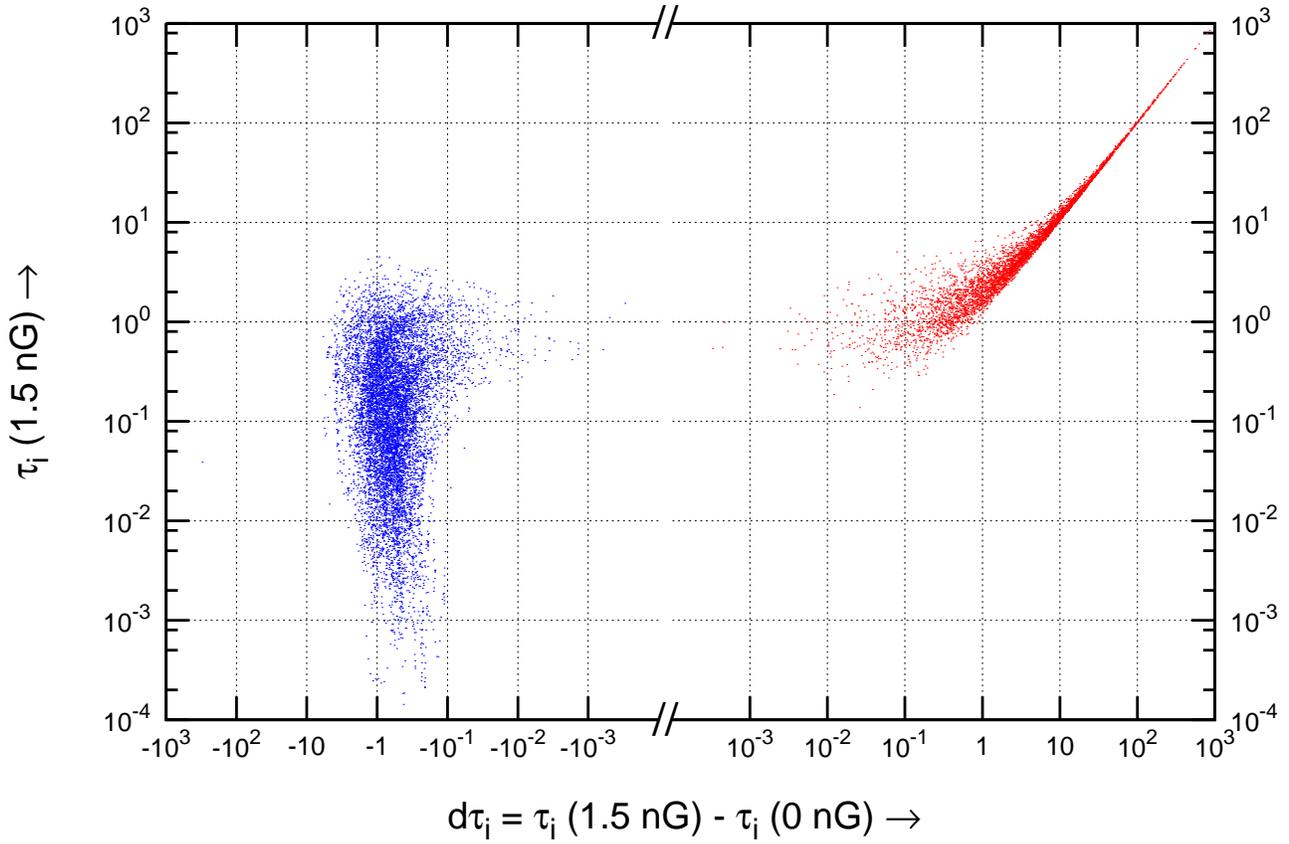}
}
\caption[]{
The distribution of $\tau_i \ (1.5 \ nG)$ versus $d\tau_i \ (= \tau_i \ (1.5 \ nG) - \tau_i \ (0 \ nG))$ 
at redshift $z = 4$.
}
\end{figure}

\vspace{0.1in}
\begin{figure}
   \centerline{
   \includegraphics[scale=0.7,angle=-90]{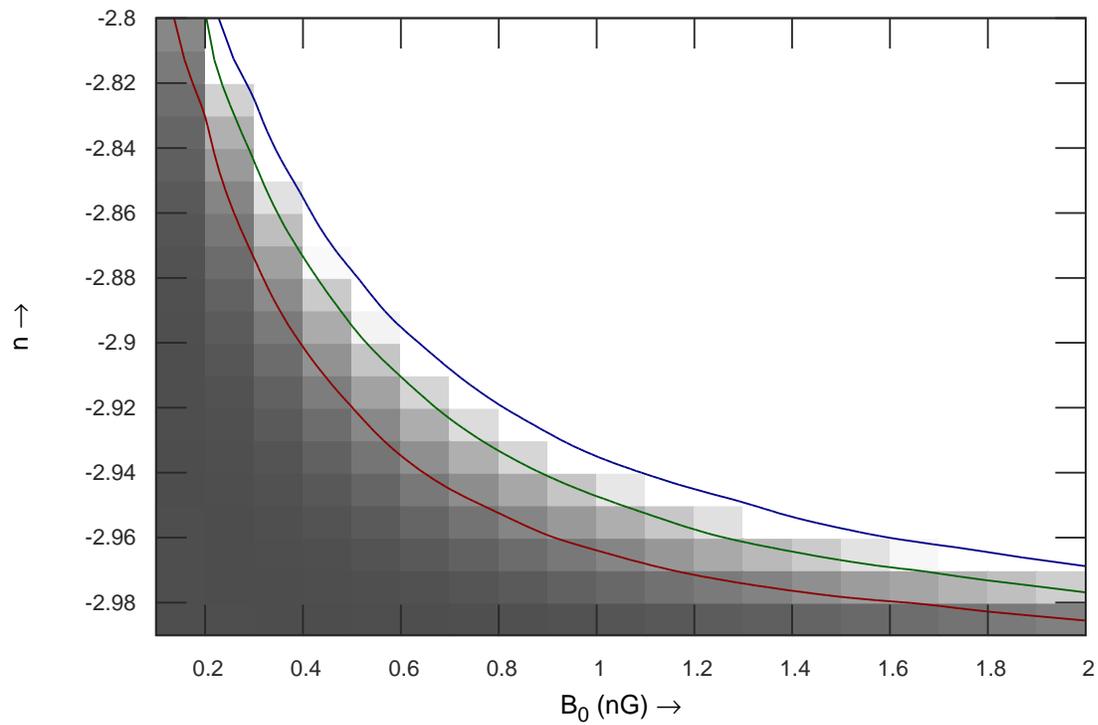}
}
\caption[]{
Allowed (the shaded) region in the $(B_0, n)$ plane, based on the $\chi^2$ analysis of $\tau_{\rm eff}$ 
against the data from \citet{FG08}. The three curves (from top to bottom) are contours at the 
5$\sigma$, 3$\sigma$ and 1$\sigma$ levels.
}
\end{figure}

\vspace{0.1in}
\begin{figure}
   \centerline{
   \includegraphics[scale=0.7,angle=-90]{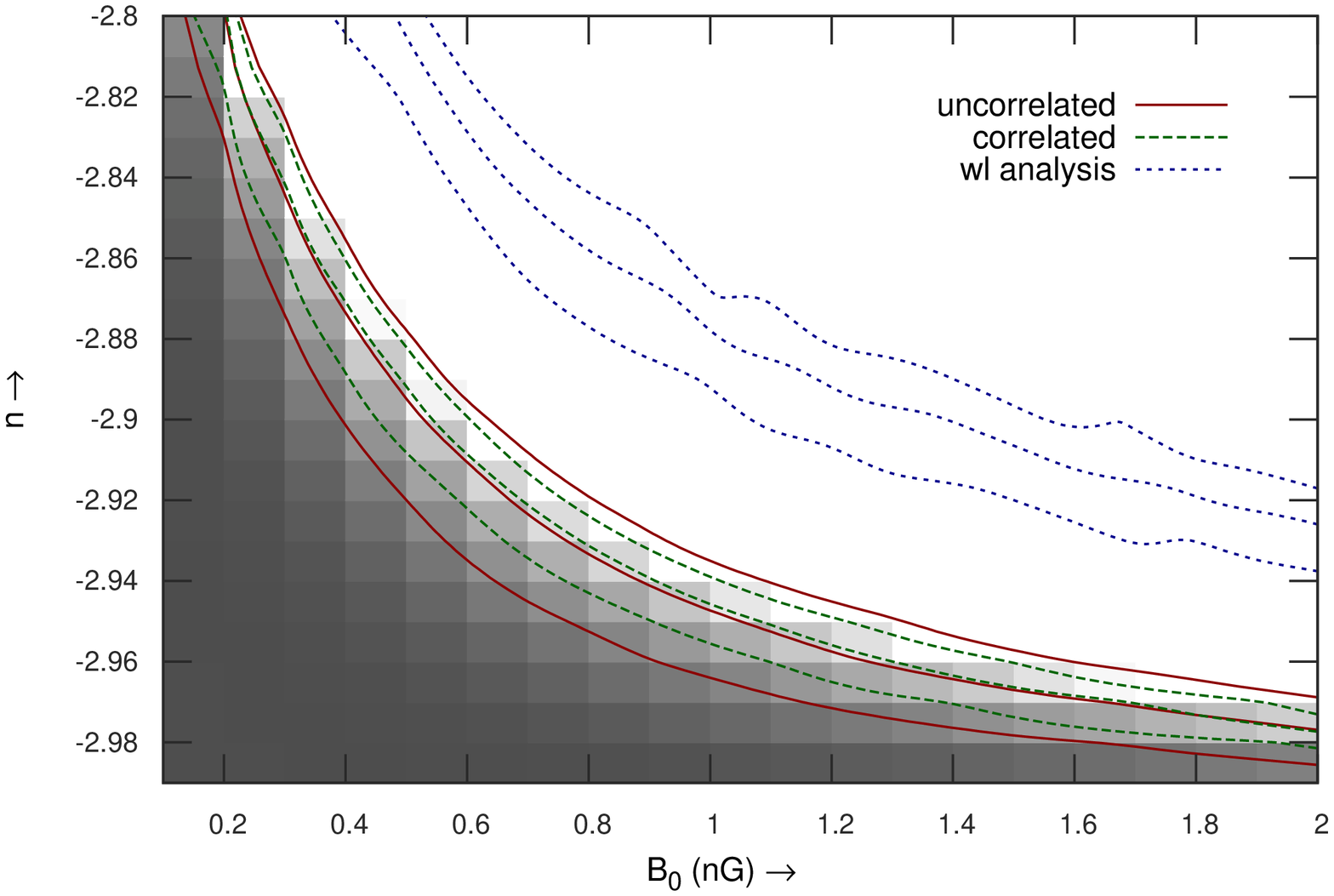}
}
\caption[]{
Contour of $\Delta\chi^2 = \chi^2_{\rm i} - \chi^2_{\rm min}$ for the present analysis (the lower triplets, solid ones correspond to 
1, 3 and 5 sigma levels for the uncorrelated case, the dashed ones are the same for the correlated case) and the previous analysis 
(the upper triplet) on ($B_0$, n) plane.
}
\end{figure}

\section{Discussion}
Primordial tangled  magnetic fields leave their   signatures on 
 cosmological observables  for  a  large range of scales from  sub-Mpc to 
$10^4 \, \rm Mpc$.  CMBR temperature and 
polarization  anisotropies  provide probes for the magnetic fields
for scales $\ga 10 \, \rm Mpc$ \citep[e.g.][]{Tina2011}. 
Recently, \citet{Yamazaki10} computed the allowed region in the $\{B_0,n\}$ plane by 
comparing the predictions of primordial magnetic field models with existing CMBR 
observations. Constraints on smaller scale come from early formation of structures induced
by PMF. The observables that impact these scales include early reionization, 
HI signal from the epoch of reionization (\citet{Sethi09,Sethi05}, \citet{SM2010}), 
cosmological weak gravitation lensing (\citet{PS12}), etc.  Other constraints
on large scale cosmological magnetic fields arise from rotation measure 
of high redshift polarized radio sources \citep[e.g.][]{1998ApJ...495..564K,2003MNRAS.342..962S,
Blasi99}; rotation measure (RM) of radio sources will be one of the methods
employed  by
 radio interferometers LOFAR and SKA to attempt to detect cosmic magnetic fields. In particular, \citet{Blasi99} considered the same physical setting (high
redshift Lyman-$\alpha$ clouds)  as
in this paper.  They computed  the RM  of Ly$\alpha$ density 
field and obtained bounds $\simeq  10^{-8} \, \rm G$
on magnetic fields with coherence length scales of the thermal Jeans length. 

 In addition to the upper bounds on the magnetic field strength obtained 
by these observables, recent results suggests that there might be 
a lower bound of $\simeq 10^{-15} \, \rm G$ on the magnetic field strength 
(e.g.  \citet{Dolag10,NV2010,Tavecchio10,Taylor11}). 
 Another lower bound is obtained from  the study of 
echo emission from the blazar Mrk 501 (\citet{Takahashi12}) which 
suggests magnetic field strength of $B_0 \gtrsim 10^{-20} G$ coherent over
 the length scale of $\sim 1$ kpc. 
This would suggest that the magnetic field strength could lie in the range  $10^{-20} < B_0 < \hbox{a few} \, 10^{-9} \, \rm G$. 
This range is still too large for a better  determination of the 
magnetic field strength.

In Figure~7, we show   the constraints from the present study 
 compare with similar constraints from cosmological gravitational 
lensing we obtained earlier (\citet{PS12}). 
In Comparison to bounds on primordial magnetic fields from 
CMBR   anisotropies (e.g. Figure~1 of \citet{Yamazaki10}),
for the entire range of spectral indices,  we obtain stronger 
 limits on $B_0$.  Other constraints 
from bispectrum and trispectrum analysis of CMBR passive scalar modes \citet{Trivedi10,Trivedi12} are
2.4 nG and 0.7 nG, they have used spectral index value n = -2.8, whereas for n = -2.8 
our analysis gives an upper bound on $B_0 \lesssim$ 0.3 nG (Figure~7).
 As noted above, these bounds are even better than 
our previous analysis with the weak-lensing data (Figure~8).

In our present analysis, we consider four parameter: $J$, $\gamma$, $B_0$, 
$n$ but no other cosmological parameters. We also do not account for 
errors arising from different realizations of the density field. Our 
current bounds can be  further improved by the inclusion of such effects. 
We note that even though the 
 magnetic field signal could be degenerate with the overall 
normalization  of the $\Lambda$CDM model as measured
 by $\sigma_8$, the current errors on the value of $\sigma_8$ 
(WMAP 7-year data give 
$\sigma_8 = 0.801 \pm 0.030$ \citep{Larson11}) are too small 
to sufficiently alter our conclusions. 

In sum: Lyman-$\alpha$ clouds provide a sensitive  probe of the matter 
power spectra at scales $\la 1 \, \rm Mpc$. Primordial magnetic field
induced matter perturbations give additional power at these scale which
can be probed using the redshift evolution of $\tau_{\rm eff}$. Our results
shows that this leads to one of the most stringent bounds on the parameters
of primordial magnetic fields. These bounds can be further improved 
by more  data on the evolution of $\tau_{\rm eff}$ at low redshifts and also
more precise data at higher redshifts. Recently,  \citet{Becker12} 
have provided a  measurement of the evolution  $\tau_{\rm eff}$ which 
is in agreement with  the data we have used (\cite{FG08}) for our 
analysis but claims better  precision. In future, similar analyses 
with such  data can give even more stringent constraints
 on the parameters  of primordial magnetic fields.

\section*{Acknowledgement}
We thank R Srianand and S Sridhar on many useful discussions on many 
aspects of this paper.

\end{document}